# Determination of the magnetic moment of a magnet by letting it fall through a conducting pipe

Sanjoy Kumar Pal[1], Soumen Sarkar[2], and Pradipta Panchadhyayee[3,4*]

[1]Anandapur H.S. School, Anandapur, Paschim Medinipur, West Bengal, India
[2]Karui P.C. High School, Hooghly, West Bengal, India
[3]Department of Physics (UG & PG), Prabhat Kumar College, Contai, Purba Medinipur, India
[4]Institute of Astronomy, Space and Earth Science, Kolkata -700054, W. B., India
[*]E-mail: ppcontai@gmail.com

## Abstract

*A novel method is proposed to determine the magnetic moment of a magnet by studying its free-falling motion inside a non-ferromagnetic and conducting pipe. The dynamics of a neodymium magnet falling inside a pipe is tracked by using sound waves of a fixed frequency generated by one smartphone and detecting acoustic resonance in the pipe simultaneously by the other. This tracking technique leads to the measurement of the terminal velocity of the falling magnet, as the interaction between the magnet and the conducting pipe creates viscosity artificially. The result obtained is verified by studying torsional oscillations of the suspended magnet and conforms to the reported value in such a low-cost setup. The experiment is designed with concepts integrating the domains of general physics, electromagnetic induction, and acoustics.*

## Introduction

Over the last decade, a significant amount of published work has covered various fields of physics like the motion of objects [1-2], general physics [3-5], optics [6-8], acoustics [9-11], magnetism [12-14], etc. using smartphone sensors. The application of smartphone sensors has gained momentum in their usage in designing innovative experiments at home desks. The wide use of smartphone sensors in physics education has opened up new possibilities for engaging students and exploring various concepts practically and interactively. For example, to teach concepts related to electromagnetic induction, Newton's third law, and Lenz's law, the use of strong neodymium magnets, along with smartphones, in classroom demonstrations [15-24] have been made very effectively. The slow motion of a falling magnet through a conducting pipe is studied to illustrate these principles. Especially, Lenz's law, which can be challenging for some students to grasp, is explored in greater detail using this demonstration by several authors. They focus on the role of eddy currents in slowing down the magnet's fall inside the conducting pipe [18-24]. Some other researchers have showcased the method for measuring the terminal velocity of the falling magnet [15-17,24] and conducted detailed analyses of electromagnetic damping [25-29].

Motivated by the above works [23,24], we have utilized acoustic resonance in an air column to monitor the free-falling motion of a body inside a conducting pipe using smartphones and measured the terminal velocity as a result of equilibrium established between the electromagnetic damping force and the weight of the magnet. Finally, we have computed the magnetic moment of the magnet under consideration from a theoretical relation involving different parameters of the pipe-magnet system and the terminal velocity of the magnet. To show the validity of the computed value of the magnetic moment of a specific magnet we employ a standard method of determining the magnetic moment by studying the torsional oscillations of the magnet.

## Experimental setup with theoretical framework

When a magnet, of mass $M$, falls through a non-ferromagnetic conducting pipe, an electromagnetic damping force acts on it and opposes its motion against gravity. The instantaneous value of the damping force ($F_{em}$) is proportional to that of the instantaneous velocity ($v$) which can be expressed as (see [30] or Appendix I[1])

$$F_{em} = \left(\frac{15}{1024}\right)\mu_0^2 m^2 \sigma \left(\frac{1}{a^3} - \frac{1}{b^3}\right) v = kv, \qquad (1)$$

where $m$ is the magnetic dipole moment of the magnet, and $\mu_0$ is the free space permittivity. The conductivity of the conducting pipe is $\sigma$ while the inner and outer radii of the pipe are $a$ and $b$, in that order. We represent the proportionality constant by $k$.

[1] The expression derived in Appendix I is valid for a small magnet, so one may think that using this expression in our working formula is questionable. However, we have found that the value of the magnetic dipole moment as obtained by this method aggress very well with that determined by the torsional oscillation method (Method II) for the same magnet. So, the use of this particular equation is justified.

After a specific time, the damping force ($F_{em}$) becomes equal to its weight and, at that condition, no net force acts on the falling magnet. So, we note that $F_{em}$ is proportional to the terminal velocity ($v_T$) attained by the magnet. Neglecting the viscous force, the buoyant force, and hydrodynamic resistance ($\propto v^2$) we can write

$$\left(\frac{15}{1024}\right)\mu_0^2 m^2 \sigma \left(\frac{1}{a^3}-\frac{1}{b^3}\right) v_T = k v_T = Mg, \quad (2)$$

where $M$ is the mass of the magnet and $g$, the acceleration due to gravity.

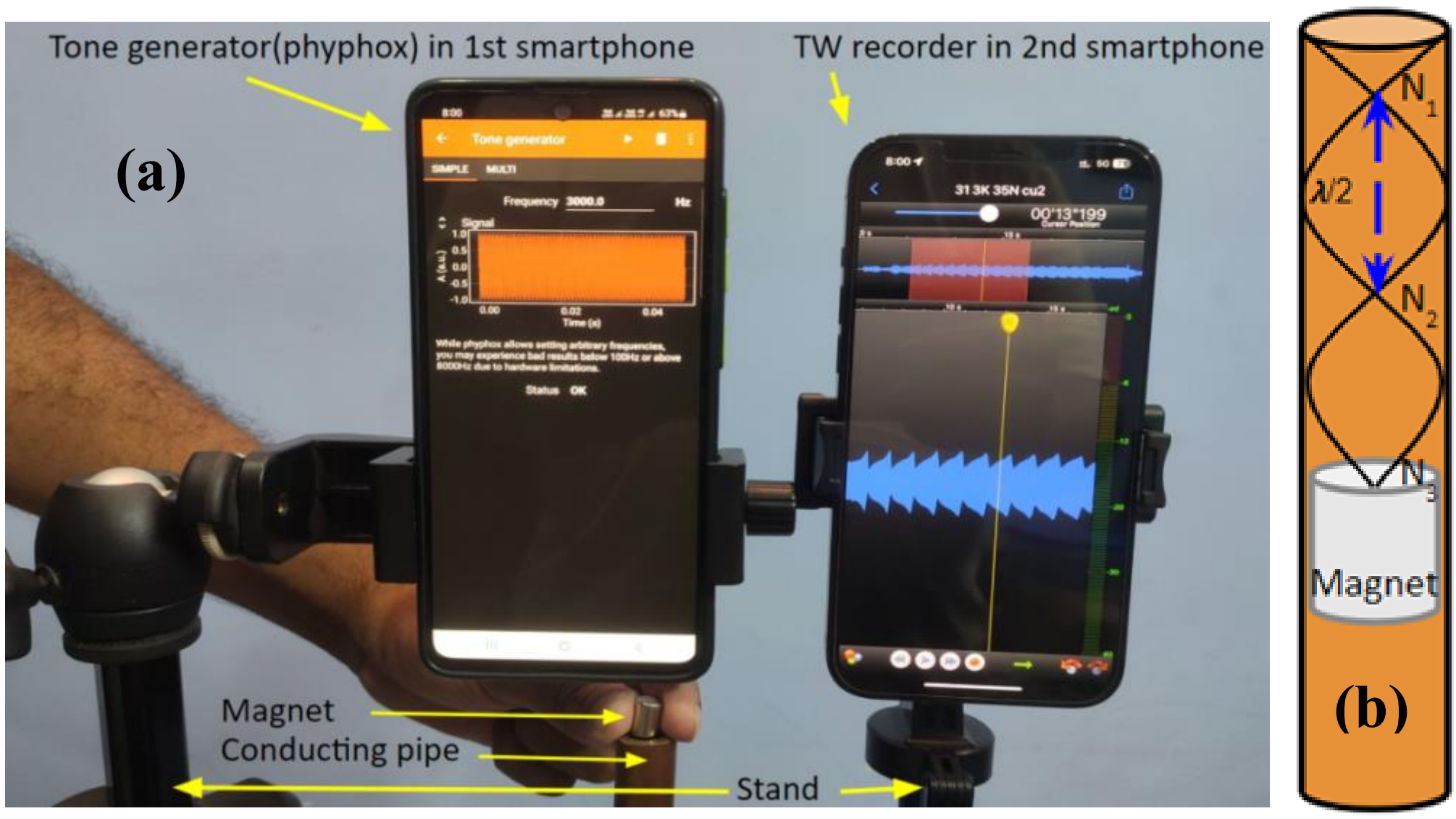

Fig. 1: (a) - Experimental setup. (b) - Generation of standing wave with nodes.

At first, a cylindrical neodymium magnet (N52) and a hollow copper pipe are used for the experiment as shown in Fig. 1a. In the case of falling of the magnet inside the copper pipe held vertically, we may consider the pipe-magnet system as a closed organ pipe. Through the constant tracking of the fall of the magnet, our first target is to measure the terminal velocity of the magnet in the 'no-net-force' condition, as seen in Eq. 2.

For the measurement, two smartphones are used, one for generating a single-frequency sound and the other for recording the admixture of the original sound and the sound produced by the pipe-magnet system. As the initial step, we hold the pipe vertically and keep it fixed by a plastic clamp. We generate a sound with an android phone (Redmi K20 pro), referred to as the first smartphone, with the Phyphox application [31], and placed at the top of the copper pipe. In particular, a pure sine wave of frequency 3 kHz is generated by the key tone generator, an in-built menu of the Phyphox application. The magnet is held just below the first smartphone by two fingers and allowed to fall. A standing wave is generated for a particular position of the falling magnet at any instant (see Fig. 1b). Considering the falling motion of the magnet inside the vertical copper pipe, the periodic variation in sound amplitude (nodes) is clearly audible, and the corresponding variations are registered via a recording by the second smartphone (Fig. 2). With this aim, we have navigated to the application TW Recorder on iPhone 12 pro max which is kept close to the open end of the pipe. We have set the frequency sampling rate at 44100 Hz for recording. The same procedure is followed for another sine wave of frequency 4 kHz and the results are verified against those for the wave of frequency 4 kHz. The whole system is just like an open-end air column with the occurrence of resonances at magnet displacements $L_n$, which can be expressed by the equation

$$L_n + \Delta l = \lambda \frac{(2n+1)}{4}; \tag{3}$$

where $n = 0, 1, 2, 3......$

Here $\Delta l$ is the end-correction term and $\lambda$ is the wavelength of the sound wave used. When the magnet completes successive displacements by an amount $\frac{\lambda}{2}$, it satisfies the resonance condition (See Fig. 2). In the condition of movement by the magnet at the terminal velocity $(v_T)$ the following expression must be satisfied:

$$L_n - L_0 = \frac{n\lambda}{2} = v_T(t_n - t_0)$$

$$\text{or}, n = \left(\frac{2v_T}{\lambda}\right)(t_n - t_0) = \left(\frac{2v_T}{\lambda}\right)\Delta t_n$$

$$\therefore \Delta t_n = \left(\frac{\lambda}{2v_T}\right) n \tag{4}$$

where $L_n(L_0)$ is the length traversed by the magnet for the $n$th (zeroth) node and $t_n(t_0)$ is the corresponding elapsed time. Any node can be considered as the zeroth node just after the instant when the magnet attains the terminal velocity. Thus, measuring the time intervals $(t_n - t_0 = \Delta t_n)$ between $n$ number of nodes, we can determine the terminal velocity $(v_T)$ of the magnet from the slope of the $n$ versus $\Delta t_n$ graph. We can also compute the value of the proportionality constant $(k)$ from the relation, $kv_T = Mg$, in Eq. (2). As is evident from the Eq. (2),

$$k = \left(\frac{15}{1024}\right)\mu_0^2 m^2 \sigma \left(\frac{1}{a^3} - \frac{1}{b^3}\right). \tag{5}$$

Putting the values of $a$, $b$, $\sigma$, $k$, and $\mu_0$, the value of $m$ is easily calculated.

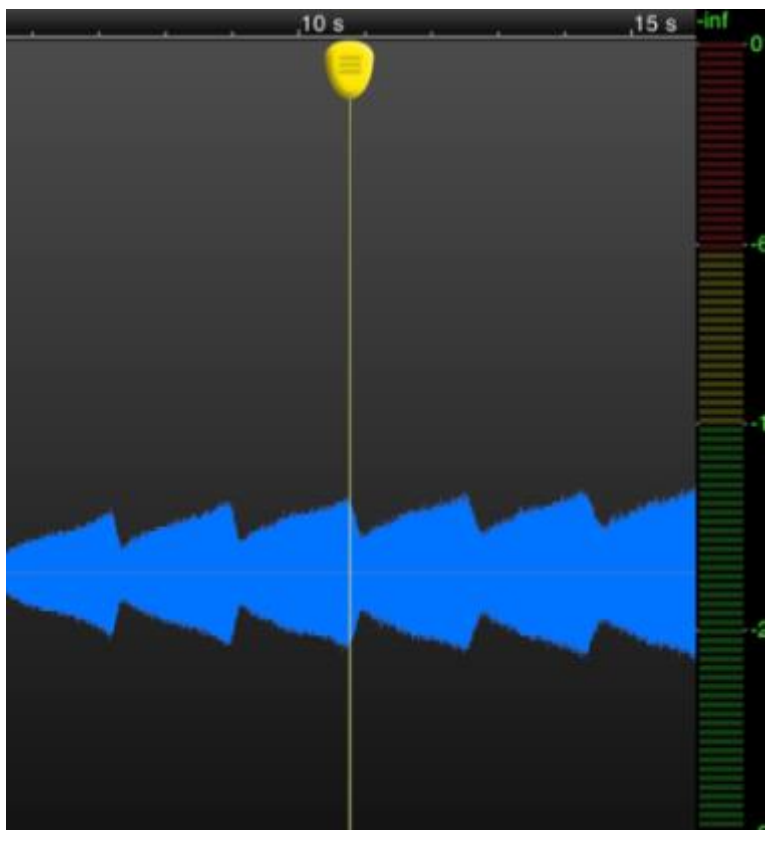

Fig. 2: The variation in the sound amplitude with time recorded by the TW Recorder application.

An aluminium pipe is also employed to establish the validity of the experimental method. The experiment is repeated with the aluminium pipe and the same N52 grade neodymium magnet. For better comprehension and to check the accuracy of the results, one comparatively weak neodymium magnets of same grade (N35) are also taken for experimentation for the copper and aluminium pipes.

To test the robustness of the method we have introduced a standard method which deals with the study of torsional oscillations of the magnet in the suspended conditions (See Fig. 3a). Later, this procedure will be referred as the second method. A specific neodymium cylindrical magnet is taken as the bob of a torsional pendulum that also uses a thread. It is held stable along the north-south direction. We allow the magnet to rotate in the *x-y* plane and measure the time period of the torsional pendulum by using the Phyphox application installed on a smartphone. The smartphone is placed at an appreciable distance (20 cm in our case) from the magnet. The horizontal component of earth's magnetic field is determined by utilising the magnetic field sensor of the smartphone via the Phyphox application. A substantial care has been taken to avoid the presence of any kind of magnetic substance near the smartphone as well as the magnet. We have noted the generated waveforms in the Phyphox window, measured the time duration for a number of oscillations, and hence the time period of the torsional pendulum (See Fig. 3b). To determine the magnetic moment of the magnet we have used the formula neglecting the torsional rigidity of the thread, as given below

$$T = 2\pi\sqrt{\frac{I}{mB_H}}, \quad (6)$$

where $T$ is the time period of torsional oscillations, $B_H$ is the horizontal component of Earth's magnetic field and $m$ is the magnetic moment of the magnet. Here, $I$ is the moment of inertia of the cylindrical magnet along the axis perpendicular to its own axis passing through its centre of mass and it can be expressed as

$$I = \frac{ML^2}{12} + \frac{Mr^2}{4}, \quad (7)$$

where $M$, $L$, and $r$ are the mass, length, and radius of the cylindrical magnet, respectively.

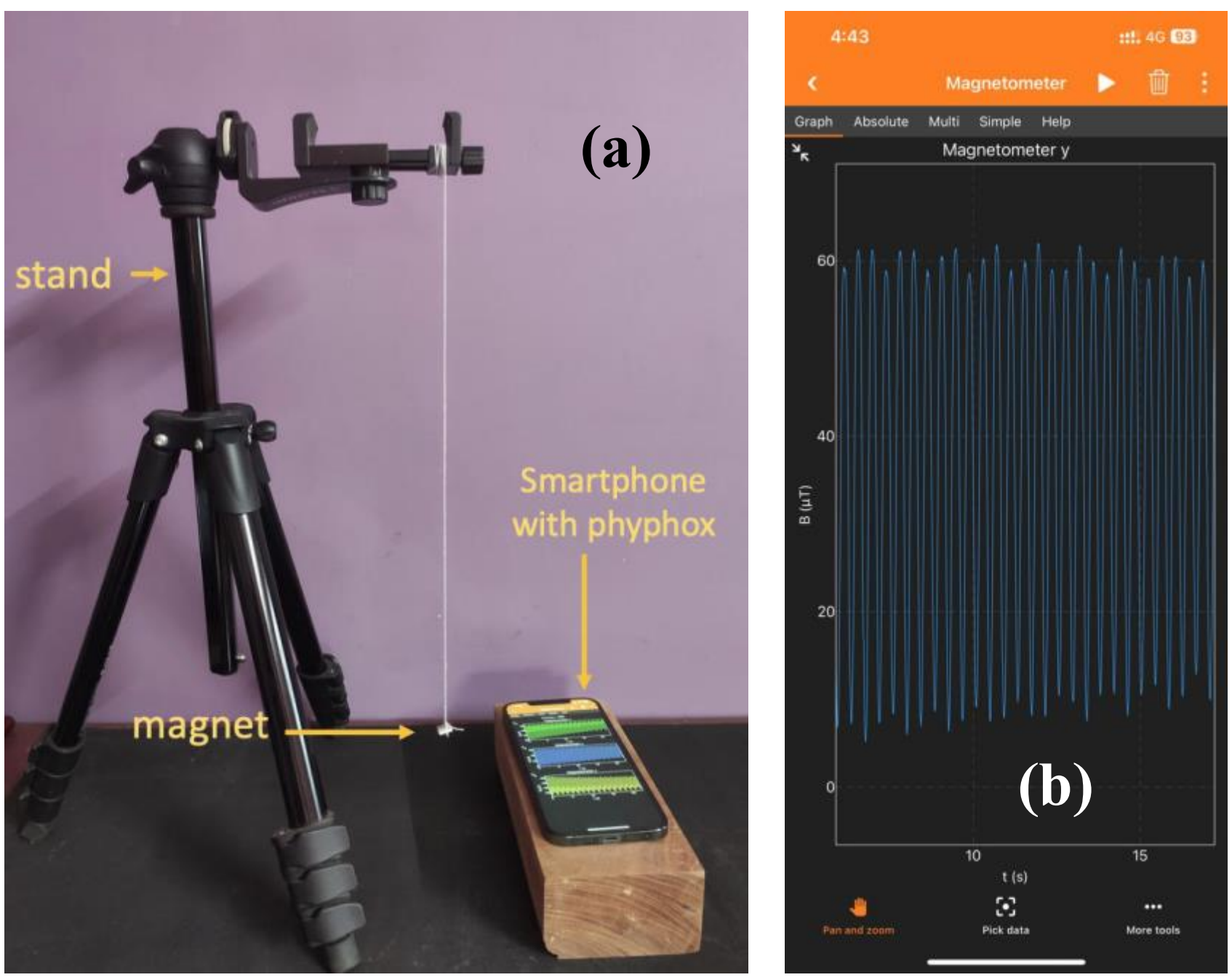

Fig. 3: (a) - Data recording of the torsional oscillations of the magnet by the Phyphox application; (b) - Measurement of the time period of oscillations from the generated waveforms in the Phyphox magnetometer application.

To measure $B_H$, the smartphone is held in the North-South direction parallel to the ground and the Phyphox magnetometer reading shows the $B_y$ value as 35 μT keeping the $B_x$ value close to zero. The direction of the smartphone is changed by 90 degree i.e., held along the East-West direction parallel to the ground. The magnetometer reading exhibits the opposite readings for the $B_x$ and $B_y$ values.

## Results and discussion

Following the first procedure above described, we have performed the experiment with a Neodymium Iron Boron magnet (N52 grade) using the copper and the aluminium pipes. The $n$ - $\Delta t_n$ graphs are plotted on the basis of these data. We have applied two wave-frequencies to establish the novelty of the method. The same experiment is repeated for a N35 grade Neodymium Iron Boron magnet using the same copper and aluminium pipes. The same procedure is followed to obtain the values of terminal velocities, and the corresponding $k$ values. Using the values of conductivity of the metals and other required values we have computed the values of magnetic moment of the N52 and N35 magnets. All the magnitudes of the physical parameters of the pipes and the magnets are clearly mentioned in the following table (Table: 1). Diameters of the magnets and the pipes and also the lengths of the magnets are measured by a digital slide callipers.

Table 1: Table for the parameters of the pipes and the magnets

Hollow Copper pipe: Length - 101.0 cm and inner (outer) diameter - 9.56 (15.86) mm

Hollow Aluminium pipe: Length - 68.0 cm and inner (outer) diameter - 12.03 (19.40) mm

Temperature - 30.0°C

Sine wave frequencies - 3 kHz and 4 kHz; Respective wavelengths at 30.0°C**:** 0.116m and 0.087m

Conductivity ($\sigma$) of copper at 30.0°C - 5.5619 x $10^7$ $(\Omega m)^{-1}$

Conductivity ($\sigma$) of aluminium at 30.0°C - 3.6354 x $10^7$ $(\Omega m)^{-1}$

| Specification of magnets | Used for copper and aluminium pipes | | |
|---|---|---|---|
| | Mass (g) | Diameter (mm) | Length (mm) |
| N52 | 3.74 | 7.98 | 9.97 |
| N35 | 3.74 | 7.98 | 9.97 |

In the case of the copper and the aluminium pipes, the node ($n$) – elapsed time ($\Delta t_n$) plots for the Neodymium Iron Boron magnets (N52 and N35 grades) are presented in Fig. 4 and 5. The corresponding graphs of $n$ - $\Delta t_n$ are linear, whose slopes for the copper and the aluminium pipes for the two wave - frequencies (3 kHz and 4 kHz) are given in Table 2. From these magnitudes of the slopes, we have calculated the values of terminal velocities of the N52 and the N35 magnets for copper and aluminium pipes following Eq. (4). Based on the Eq. (2), we have computed the respective values of $k$. The value of magnetic moment of the N52 magnet is then determined putting the value of $k$ and the corresponding conductivities ($\sigma$) of copper and aluminium.

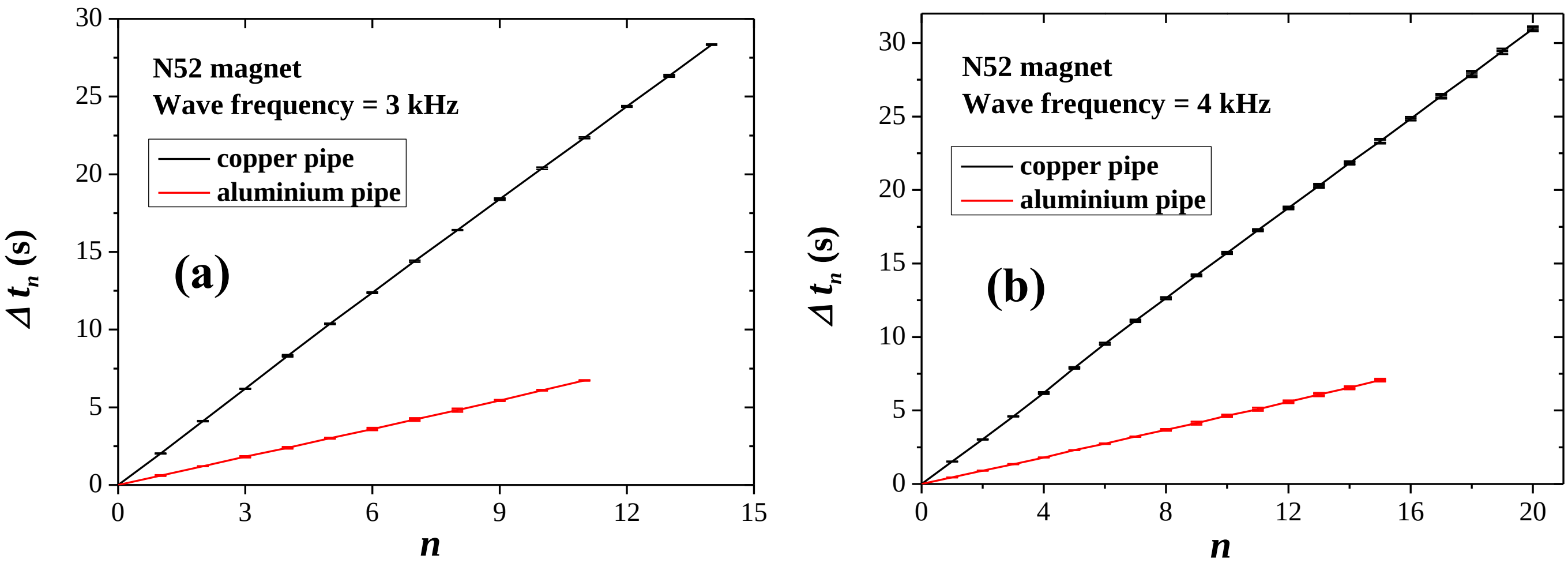

Fig. 4: Plot (with error bars) of the node ($n$) against the elapsed time ($\Delta t_n$) for the N52 magnet with $n = 0$ as reference, when falling inside the copper/aluminium pipe with the applied wave frequency (a) 3 kHz, and (b) 4 kHz.

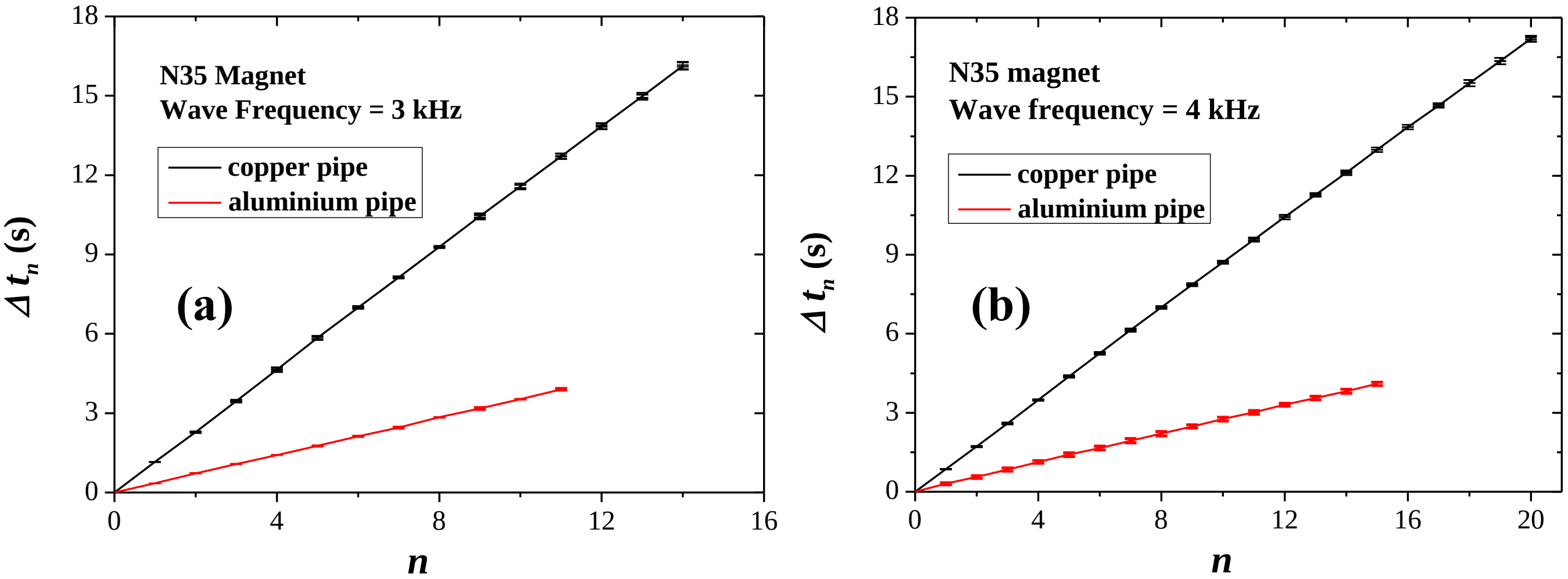

Fig. 5: Plot (with error bars) of the node ($n$) against the elapsed time ($\Delta t_n$) for the N35 magnet with $n = 0$ as reference, when falling inside the copper/aluminium pipe with the applied wave frequency (a) 3 kHz, and (b) 4 kHz.

Table 2: Table for the physical quantities in the case of the N52 and the N35 magnets (Method I)

| Magnet Grade | Material | Frequency (kHz) | Slope (s) | Terminal Velocity ($ms^{-1}$) | $k$ ($kg\ s^{-1}$) | Magnetic Moment (m) $A\text{-}m^2$ | Mean Magnetic Moment (m) $A\text{-}m^2$ |
|---|---|---|---|---|---|---|---|
| N52 | copper | 3 | 2.0190 | 0.0287 | 1.27716 | 0.37 | 0.37 ± 0.01 |
| | | | 2.0243 | 0.0287 | 1.28052 | 0.37 | |
| | | | 2.0272 | 0.0286 | 1.28234 | 0.37 | |
| | | 4 | 1.5538 | 0.0280 | 1.31049 | 0.38 | |
| | | | 1.5708 | 0.0277 | 1.32490 | 0.38 | |
| | | | 1.5755 | 0.0276 | 1.32887 | 0.38 | |
| | aluminium | 3 | 0.6138 | 0.0945 | 0.38825 | 0.36 | |
| | | | 0.6167 | 0.0941 | 0.39009 | 0.36 | |
| | | | 0.6104 | 0.0950 | 0.38612 | 0.36 | |
| | | 4 | 0.4813 | 0.0904 | 0.40591 | 0.37 | |
| | | | 0.4655 | 0.0934 | 0.39266 | 0.37 | |
| | | | 0.4635 | 0.0939 | 0.39093 | 0.36 | |
| N35 | copper | 3 | 1.1632 | 0.0499 | 0.73581 | 0.28 | 0.28 ± 0.01 |
| | | | 1.1621 | 0.0499 | 0.73513 | 0.28 | |
| | | | 1.1511 | 0.0504 | 0.72819 | 0.28 | |
| | | 4 | 0.8669 | 0.0502 | 0.73113 | 0.28 | |
| | | | 0.8601 | 0.0506 | 0.72541 | 0.28 | |
| | | | 0.8566 | 0.0508 | 0.72249 | 028 | |
| | aluminium | 3 | 0.3575 | 0.1622 | 0.22616 | 0.28 | |
| | | | 0.3540 | 0.1639 | 0.22390 | 0.28 | |
| | | | 0.3486 | 0.1664 | 0.22049 | 0.27 | |
| | | 4 | 0.2701 | 0.1611 | 0.22781 | 0.28 | |
| | | | 0.2708 | 0.1523 | 0.23658 | 0.28 | |
| | | | 0.2753 | 0.1551 | 0.23221 | 0.28 | |

As mentioned above regarding the measurement of the magnetic moments of the magnets under consideration, we have also followed a second method. The physical constants like masses, lengths and radii of the N52 and N35 magnets are tabulated below in Table 3 along with the measurement of the time periods of torsional oscillations of the magnets and finally the computation of the magnetic moments.

Table 3: Table for the physical quantities in the case of the N52 and the N35 magnets (Method II)

| Magnet Grade | $B_H$ (μT) | Number of Oscillations | Time taken (s) | Time period, $T$ (s) | Mag. Moment, $m$ (A-m$^2$) | Mean Mag. Moment, $m$ (A-m$^2$) |
|---|---|---|---|---|---|---|
| N52 | 35 | 68 | 25 | 0.3676 | 0.38 | 0.37 ± 0.01 |
| | | 78 | 29 | 0.3718 | 0.37 | |
| | | 61 | 23 | 0.3770 | 0.36 | |
| | | 51 | 19 | 0.3725 | 0.37 | |
| | | 73 | 27 | 0.3699 | 0.38 | |
| N35 | 35 | 46 | 20 | 0.4348 | 0.27 | 0.28 ± 0.01 |
| | | 61 | 26 | 0.4262 | 0.28 | |
| | | 58 | 25 | 0.4310 | 0.28 | |
| | | 72 | 31 | 0.4306 | 0.28 | |
| | | 58 | 25 | 0.4310 | 0.28 | |

To sum up the computed results, it is evident from Table 2 and Table 3 that the values of the magnetic moments of the N52 and N35 grade magnets determined by the two methods are very close to each other. The average magnetic moment of the N52 neodymium magnet is 0.37 ± 0.01 A-m$^2$, whereas we have obtained the average value of the magnetic moment of the N35 magnet as 0.28 ± 0.01 A-m$^2$. We observe that, for both input frequencies, the value of the magnetic moment of a specific magnet (N52/N35) computed for the copper pipe is a little bit less than that calculated for the aluminium pipe. This deviation may be attributed to the greater thickness and larger air gap between the magnet and the aluminium pipe compared to the same factors in the case of the copper pipe.

## Conclusion:

To conclude, a theoretical model is developed for determining the magnetic moment of a magnet by tracking the movement of a free-falling neodymium magnet through a conducting pipe. An interesting experiment is designed implementing the associated theoretical ideas on general physics, electromagnetic induction, and acoustics. By utilizing two smartphones, we have analysed the detected acoustic resonances and measured the terminal velocity of the magnets and determined their magnetic moments. To establish the validity of the result, we have performed the experiment with two types of the Neodymium Iron Boron magnet (grades: N52 and N35) and two metal pipes of copper and aluminium for two different frequencies. For comparison of the magnetic moments of the two magnets, we have employed another standard method and finally obtained very close values of the magnetic moments by the two methods. The proposed method is effective in accurately determining the magnetic moment of a magnet.

## Appendix I:

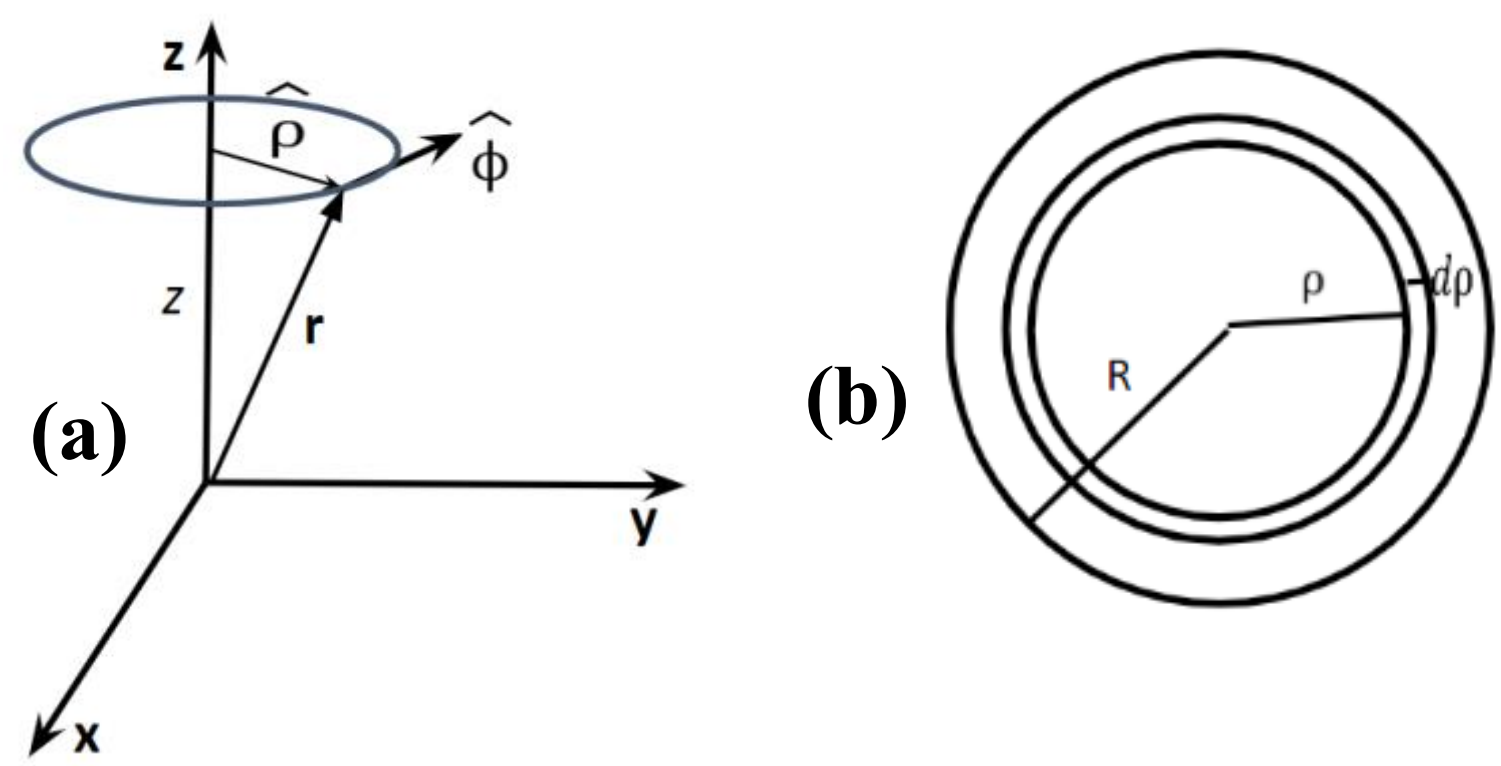

Fig. A1: (a) - The ring structure specified in the Cartesian and the Cylindrical coordinate systems; (b) – A planar view of the ring in the $\rho$-$\varphi$ plane.

This section presents the detailed calculation of the force exerted by the conducting pipe on a magnet falling inside it. An electromotive force (EMF) is induced in the conducting pipe due to the falling motion of the magnet. As a natural consequence, an opposing force obeying the Lenz's law is expected to be produced by the pipe on the falling magnet. For the sake of simplicity, we can assume that the pipe is made of very thin rings. We can find the force exerted by a ring and, finally summing the force of each ring, the force produced by the pipe. For better comprehension, the total calculation is split into the three parts:

i. Dipole Approximation

ii. EMF by using Faraday's Law

iii. Force due to the pipe

**i. Dipole Approximation**

In the magnetic dipole approximation, the magnetic moment $m$ of the falling magnet is assumed to be a single magnetic dipole of moment $m$ oriented along the positive $z$-axis; its size is supposed to be small compared to the radius of a ring of the pipe. The well-known expression of the magnetic field ($\vec{B}$) due to the dipole placed along the $z$-direction is given by:

$$\vec{B} = \frac{\mu_0}{4\pi}\left[\frac{3(\vec{m}.\vec{r})\vec{r}}{r^5} - \frac{\vec{m}}{r^3}\right] \tag{A1}$$

It is more useful to choose a cylindrical coordinate system [Fig. A1] where $\vec{B}$ can be represented by $\vec{B} = B_\rho\hat{\rho} + B_z\hat{z} + B_\varphi\hat{\varphi}$. For the cylindrical symmetry, $B_\varphi\hat{\varphi} = 0$.

$$\text{So, } B_\rho = \sqrt{{B_x}^2 + {B_y}^2} = \frac{\mu_0\, m}{4\pi}\frac{3z}{r^5}\sqrt{x^2+y^2} = \frac{\mu_0}{4\pi}\frac{3z\rho}{(\rho^2+z^2)^{5/2}}, \tag{A2}$$

$$\text{and } B_z = \frac{\mu_0 m}{4\pi}\left[\frac{3z^2}{(\rho^2+z^2)^{5/2}} - \frac{1}{(\rho^2+z^2)^{3/2}}\right]. \tag{A3}$$

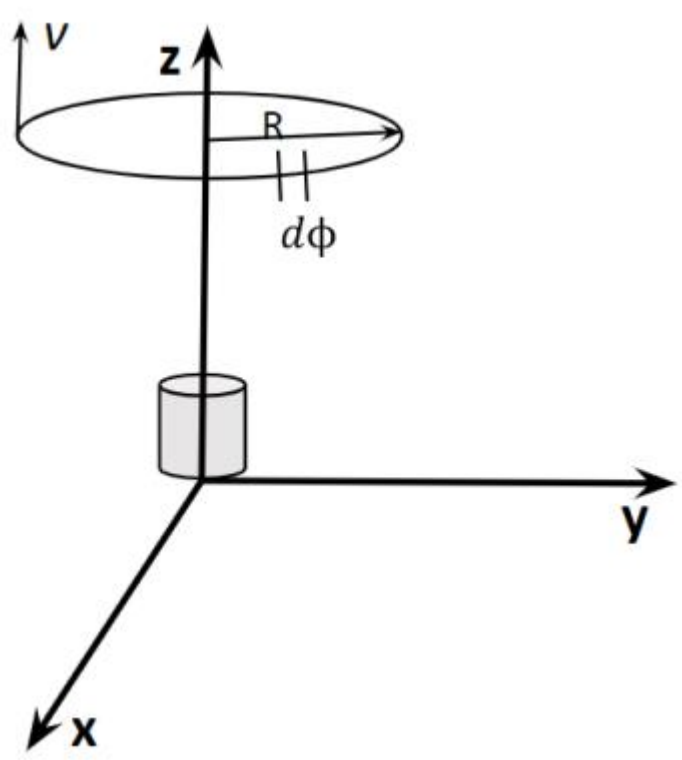

Fig. A2: The pipe-magnet system specified in the cylindrical coordinate system.

**ii. EMF – Faraday's Law:**

Here we have used Faraday's law: $\varepsilon = -\frac{d\varphi_B}{dt}$, (A4)

where $\varepsilon$ is the induced EMF in the ring and$\varphi_B$, the magnetic flux. We can compute the flux as

$$\varphi_B = \int B_z\, dA = \int_0^R B_z\, 2\pi\rho d\rho$$

$$= \frac{\mu_0 m}{4\pi}\int_0^R \left[\frac{3z^2}{(\rho^2+z^2)^{5/2}} - \frac{1}{(\rho^2+z^2)^{3/2}}\right] 2\pi\rho d\rho = \frac{\mu_0 m}{2}\frac{R^2}{(R^2+z^2)^{3/2}}.$$

Putting the value of $\varphi_B$in Eq. (A4) we obtain the expression of the induced EMF,

$$\varepsilon = -\frac{d\varphi_B}{dt} = -\frac{d\varphi_B}{dz}\frac{dz}{dt} = -\frac{d\varphi_B}{dz}v = \frac{3\mu_0 m}{2}\frac{R^2 z}{(R^2+z^2)^{5/2}}v \tag{A5}$$

The above expression of the EMF can also be verified from the expression of the work done by the Lorentz force.

**iii. Force due to the pipe:**

The magnet is assumed to reside at the origin and is aligned along the positive $z$-direction. On the contrary, it is considered that the ring is moving with the velocity $v$ along the positive z-direction. From the third law of Newton's laws of motion, we know that the force exerted on the magnet due to the ring is equal and opposite to that acted on the ring due to the magnet. For the cylindrical symmetry, the net force in the positive *z*-direction becomes:

$$F_z^{mag} = \int dF_z^{ring} = -\int I(\vec{dl} \times \vec{B})_Z = \int IB_\rho \, dl = \int_0^{2\pi} I \, B_\rho R d\varphi = 2\pi RIB_\rho \, , \qquad \text{(A6)}$$

where *I* is the induced current on the ring.

The induced current is found out from Ohm's law: $I = \frac{V}{r} = VC = \frac{V\sigma A}{L}$, where $C$ is the conductance, $\sigma$ is the conductivity of the material, *A* is the cross-sectional area and *L* is the length.

If we divide the pipe to infinitesimally thin rings, where each infinitesimal ring has a cross sectional area of *dA* and an induced current *dI*, each part exerts an infinitesimally small element of force *dF*, From Eq. (A6) we write: $dF = 2\pi RB_\rho dI = 2\pi RB_\rho \left(\frac{\varepsilon\sigma dA}{2\pi R}\right) = B_\rho \sigma\varepsilon dA$ (A7)

As $\varepsilon = 2\pi Rv B_\rho$ and $B_\rho = \frac{3\mu_0 m}{4\pi} \frac{R_z}{(R^2+z^2)^{5/2}}$, Eq. (A7) takes the form:

$$dF = \frac{9\mu_0 m^2}{8\pi} v\sigma \frac{R^3 z^2}{(R^2+Z^2)^5} dzdR \qquad \text{(A8)}$$

To calculate the total force we integrate Eq. (A8) within the limits of the inner radius *a* and the outer radius *b*.

$$F = \int dF = \frac{9\mu_0^2 m}{8\pi} v\sigma \int_a^b \int_{-\infty}^{\infty} \frac{R^3 z^2}{(R^2+z^2)^5} dzdR \qquad \text{(A9)}$$

As the pipe is infinitely long, the range of $Z$ is taken from $-\infty$ to $\infty$.

For simplifying the integration in Eq. (A9) we make the trigonometric substitution, $z = R\, tan\theta$. Eq. (A9) can be written as: $F = \frac{9\mu_0^2 m}{8\pi} v\sigma \int_a^b \alpha\,(\theta)\, dR$, (A10) where $\alpha\,(\theta) = \int_{-\frac{\pi}{2}}^{\frac{\pi}{2}} \frac{R^5\, tan^2\theta\, R\, sec^2\theta}{R^{10}\, sec^{10}\theta}\, d\theta.$ (A11)

Eq. (A11) has been simplified via the following steps:

$$\alpha\,(\theta) = \frac{1}{R^4} \int_{-\frac{\pi}{2}}^{\frac{\pi}{2}} (cos^6\theta - cos^8\,\theta)\, d\theta = \frac{1}{R^4}(F_6 - F_8); \qquad \text{(A12)}$$

where $F_{2n} = \int_{-\frac{\pi}{2}}^{\frac{\pi}{2}} cos^{2n}\,\theta\, d\theta = (2n-1)[F_{2n-2} - F_{2n}].$ (A13)

From Eq. (A13) it is clear that $F_{2n} = \frac{2n-1}{2n} F_{2n-2} = \frac{2n-1}{2n}\left(\frac{2n-3}{2n-2}\right) F_{2n-4} = \cdots$.

The generalized form becomes $F_{2n} = \frac{(2n)!}{(2^n n!)^2}\pi.$ (A14)

On the basis of Eqs. (A12) and (A14) we write the final expression as

$$F = \frac{9\mu_0^2 m^2}{8\pi} v\sigma \frac{5\pi}{128} \int_a^b \frac{dR}{R^4} = \frac{15}{1024} \mu_0^2 m^2 \sigma \left(\frac{1}{a^3} - \frac{1}{b^3}\right) v \qquad \text{(A15)}$$

## Acknowledgement:

We are thankful to Lukas Hanson and Baris Altunkaynak for the related theory, which was discussed by them inhttps://www.youtube.com/watch?v=2-iEVFICIqMand complied by us in Appendix I. We thank to Dr S. C. Samanta and Dr D. Syam for their valuable suggestions to improve the quality of the manuscript.